# Development and Interpretation of a Neural Network-Based Synthetic Radar Reflectivity Estimator Using GOES-R Satellite Observations

**Kyle A. Hilburn**
Cooperative Institute for Research in the Atmosphere

**Imme Ebert-Uphoff**
Colorado State University

**Steven D. Miller**
Cooperative Institute for Research in the Atmosphere

Corresponding author:
Kyle Hilburn
Kyle.Hilburn@colostate.edu

**Abstract**

The objective of this research is to develop techniques for assimilating GOES-R Series observations in precipitating scenes for the purpose of improving short-term convective-scale forecasts of high impact weather hazards. Whereas one approach is radiance assimilation, the information content of GOES-R radiances from its Advanced Baseline Imager (ABI) saturates in precipitating scenes, and radiance assimilation does not make use of lightning observations from the GOES Lightning Mapper (GLM). Here, a convolutional neural network (CNN) is developed to transform GOES-R radiances and lightning into synthetic radar reflectivity fields to make use of existing radar assimilation techniques. We find that the ability of CNNs to utilize spatial context is essential for this application and offers breakthrough improvement in skill compared to traditional pixel-by-pixel based approaches. To understand the improved performance, we use a novel analysis methodology that combines several techniques, each providing different insights into the network's reasoning. Channel withholding experiments and spatial information withholding experiments are used to show that the CNN achieves skill at high reflectivity values from the information content in radiance gradients and the presence of lightning. The attribution method, layer-wise relevance propagation, demonstrates that the CNN uses radiance and lightning information synergistically, where lightning helps the CNN focus on which neighboring locations are most important. Synthetic inputs are used to quantify the sensitivity to radiance gradients, showing that sharper gradients produce a stronger response in predicted reflectivity. Finally, geostationary lightning observations are found to be uniquely valuable for their ability to pinpoint locations of strong radar echoes.

# 1. Introduction

Geostationary Operational Environmental Satellite (GOES) imagery is a key element of U.S. operational weather forecasting, supporting the need for high-resolution, rapidly refreshing imagery for situational awareness (*Line et al. 2016*). While used extensively by human forecasters, its usage in data assimilation (DA) for Numerical Weather Prediction (NWP) models is limited. Instead DA makes greater usage of microwave and infrared sounder data on low Earth orbiting satellites (*Lin et al. 2017*). Sounders provide more vertically resolved information than imagers, which is advantageous for characterizing the three-dimensional model state, but are carried almost exclusively on low-earth orbiting satellites—providing global coverage but at the expense of coarse temporal resolution and latency that can reach 1.5 hr or more. Geostationary imagers provide much faster temporal refresh (now 10 minutes for Full Disk and 5 minutes over CONUS) and very low latency over a limited field of regard. Thus, there is an opportunity for operational DA to benefit from the high volume of low-latency, complementary data coming from the global constellation of geostationary imagers.

Operational DA, despite recent scientific advances (*Zhang et al. 2019*), still largely ignores satellite pixels with cloud cover or precipitation (*Zupanski et al. 2011*). This means that the most dynamic areas from the standpoint of precipitation, having significant impacts on human activities, are also the areas that have the least amount of data to constrain estimates of the current atmospheric state. There resides an opportunity for operational DA to benefit from new, computationally efficient techniques that can provide information in these cloud and precipitation affected regions.

The objective of this research is to assimilate GOES-R Series observations from its Advanced Baseline Imager (ABI; *Schmit et al., 2017*) and GOES Lightning Mapper (GLM; *Goodman et al., 2013*) in precipitating scenes for the purpose of improving short-term convective-scale forecasts of high impact weather hazards. One approach is radiance assimilation (RA), which has the advantage of being physically based, making it simpler to interpret. However, the information content of individual pixels saturates around optical depths of 160 (8) during day (night). These values roughly correspond to composite reflectivity (REFC, the vertical maximum radar reflectivity in the column) of 20-25 (0-5) dBZ. This truncated sensitivity means, in turn, that RA holds only limited information about precipitating scenes. Moreover, RA does not handle lightning information, despite the direct linkage between lightning activity and convective precipitation. Another approach is machine learning (ML), which is statistically based, making it harder to interpret. However, ML has the advantage of making use of gradient information, which we will show provides reliable information content up to REFC of about 50 dBZ. Moreover, ML provides an effective framework for using lightning information together with radiance information.

The Rapid Refresh Forecast System (RRFS) that encompasses the Rapid Refresh (RAP) and High-Resolution Rapid Refresh (HRRR) models has long used radar reflectivity to estimate latent heating in order to spin-up convection in the models. (*Benjamin et al., 2016*). Using this pathway for GOES information would require producing 3D fields of radar reflectivity. We will treat this problem as vertically separable, first estimating the spatial distribution of REFC, and then estimating the vertical profile in a second step. This paper will tackle the REFC part of the problem, and to focus on convective-scale applications, will consider warm season convection over eastern CONUS where radar coverage is best. We describe the development of a convolutional neural network (CNN) for that purpose, including architecture selection and a novel approach to design a loss function to deal with class imbalances of REFC values.

Performance is evaluated using metrics including not only the mean-square-error (MSE), but also coefficient of determination ($R^2$), categorical metrics (probability of detection, false alarm rate, critical success index, and categorical bias) at various output threshold levels, and evaluation of the root-mean-square-difference (RMSD) binned over the range of true output values. Besides producing a trained and evaluated model, we seek to understand how our network makes its predictions. This paper uses a novel analysis methodology to open the lid of the "black box" and identify the strategies the NN is using that produce good skill.

We will begin with short descriptions of the "source" observations from the GOES-R ABI (Section 2a) and GLM (Section 2b), followed by our "target" observations from the Multi-Radar Multi-Sensor MRMS (Section 2c). The approach for constructing the ML training and validation datasets is described in Section 2d. The CNN architecture is described in Section 2e, and a novel approach for constructing a weighted loss function is given in Section 2f. The resulting CNN prototype has been dubbed "GOES Radar Estimation via Machine Learning to Inform NWP" (GREMLIN). In Section 3a, we begin with an overall characterization of the performance of GREMLIN, finding remarkably good performance, even at higher REFC values. In order to explain how GREMLIN makes such predictions, in Section 3b we selectively disable specific abilities of this model, resulting in a progression of simpler models, and analyze their results. By examining the predictions from various models (withholding certain channels and/or withholding spatial information), many insights can be gleaned. To examine the use of spatial information, we discuss and visualize the Effective Receptive Field of GREMLIN (Section 3c). To understand how the network is making its predictions, and in particular how it uses radiance information and lightning together, we apply the attribution method Layer-wise Relevance Propagation (Section 3d). Finally, we construct synthetic inputs representing different meteorological scenarios to probe the network's response and gain further insights into the use of spatial information by the network and to characterize its sensitivity (Section 3e). Section 4 presents conclusions.

## 2. Data and Methodology
### a. Advanced Baseline Imager

This study is making use of radiances from the GOES-R ABI (*Schmit et al. 2017*) on GOES-16. We are taking advantage of the higher spatial resolution (2 km) and faster temporal refresh (5-min over CONUS) in this study. In order to produce a unified Day-Night algorithm, we are focusing on just infrared channels, and for maximum portability and compatibility to legacy observing systems, using the "heritage" channels:
- Channel 7, 3.9-micron, shortwave infrared window
- Channel 9, 6.9-micron, mid-level water vapor (~442 mb)
- Channel 13, 10.3-micron, clean longwave infrared window

The conversion and calibration of observed radiances *Rad* to brightness temperatures $T_B$ for GOES ABI follows *Schmit et al. (2010)*,

$$T_B = \frac{c_2}{\ln(c_1/Rad + 1)} \tag{1a}$$

$$T_{B,C} = \frac{T_B - b_1}{b_2} \tag{1b}$$

where $c_1$ and $c_2$ are the wavenumber-dependent coefficients used to compute the monochromatic $T_B$, and $b_1$ and $b_2$ are spectral bandpass correction offset and scale for calculating the calibrated brightness temperature $T_{B,C}$. These coefficients are provided in the GOES L1b netcdf data files.

We note that during the daytime, use of the optical depth information from the red band (ABI Band 2; 0.64 µm) reflectance and the cloud particle size and phase information from ABI Band 6 (near-infrared; 2.2 µm) reflectance provide significant additional skill, however, use of these bands is beyond the scope of this paper.

Two angular quantities are especially relevant to the interpretation of ABI imagery: satellite viewing zenith angle and solar zenith angle. This study makes use of GOES-16 data from 2019, positioned in its operational East position (75.2ºW). In this slot, the satellite viewing zenith angle increases from 35º in northern Florida to 60º in North Dakota. Since we are focusing over just CONUS, we can ignore viewing zenith angle dependence because the limb cooling effect (*Elmer et al. 2016*) is small in the atmospheric window bands we are considering. We will also consider an example of storms over Colorado in 2017 when GOES-16 was in its initial check-out position (89.5ºW), which had satellite viewing zenith angles around 45º, compared to 50º in the operational East position. ABI Band 7 (3.9 µm) has a daytime solar reflective component, which would argue for solar zenith angle being be part of the model, but our results suggest this has a minor impact on the CNN results. In Section 3a we consider the skill of the model as a function of the solar zenith angle, which was calculated following *NOAA NESDIS (1998)*.

In a traditional pixel-based retrieval, correcting the effect of parallax (*Vicente et al., 2002* and Appendix A in *Miller et al., 2018*) is essential for matching up satellite data with radar data on these scales. The main uncertainty with parallax correction is estimating the height of the cloud. One can assume a fixed height, such as 10 km, to substantially reduce the error, at least for the deep clouds that are most relevant; or one can use a cloud top height product, but this can introduce blank spots in the parallax corrected imagery when low and high clouds are next to each other. To remove parallax offsets to first order, we assumed a height of 10 km. Besides residual parallax errors, there are other reasons for spatial displacements, namely vertical wind shear, and the CNN seems to learn to apply additional spatial displacements on its own based on what it sees in the training data.

*b. Geostationary Lightning Mapper*

The other major advancement provided by the GOES-R Series is real-time lightning observation from the GLM (*Goodman et al. 2010, Goodman et al. 2013*). Lightning is incredibly useful in constructing synthetic radar fields because of its association with the locations of strong updrafts within an embedded convective complex. The physical basis for this association is the strong spatial relationship between lightning flash rates, updraft vertical velocity (*W*), and latent heat release. If the terminal velocity of a raindrop goes as the square-root of the diameter, then it can be shown that mass (and latent heating) goes as $W^6$ and (linear) radar reflectivity factor goes as $W^{12}$. Meanwhile, using simple electrostatic arguments (*Price and Rind 1992, Boccippio 2002*), one can derive that lightning flash rate goes as $W^5$ for continental thunderstorms.

Much of the research on using GLM for severe weather has focused on the temporal variability, in particular lightning jumps (*Schultz et al. 2009, Schultz et al. 2015*). However, temporal variability of optically sensed lightning can provide misleading signals. This seems to be due to time varying detection efficiency effects related to the production of cloud ice (*Rutledge et al. 2020*), and also possibly to the unsteady nature of updrafts. Instead spatial variability contains more reliable information content, and supplements missing information at very high optical depths, especially at night. While there is spatial variability in GLM detection efficiency (*Marchand et al. 2019*), our CNN is more sensitive to the presence of lightning rather than the magnitude of lightning activity, which makes it less sensitive to GLM detection efficiency issues.

GLM maps total lightning with a spatial resolution of 8 km at nadir to 14 km at the limb. The basic unit of data, called an "event", is a gridded quantity, integrating all lightning pulses within the grid box over a 2 ms time window. The Lightning Cluster and Filter Algorithm (LCFA) combines adjacent lightning pixels into "groups", which are then clustered into "flashes" using a 330 ms temporal window and a 16.5 km spatial window. Thus, groups and flashes are represented as point observations consisting of a latitude, longitude, time, and area. The LCFA also performs filtering to reduce false alarms. Examination of a few sample storms found the best results (in terms of correlation with REFC) occur when using GLM groups, because they provide more "filled in" maps than using flashes. For this work we create group-extent density maps using the group area, assuming it is circular, and accumulating data over 15-minute intervals. We tested 5-minute accumulation periods but found this finer temporal granularity produced stratiform areas that flicker on and off from frame to frame. These lighting data units are given as: groups 5-min$^{-1}$ km$^{-2}$.

*c. Multi-Radar Multi-Sensor Dataset*

The target dataset to which we are training is the quality-controlled composite reflectivity from the Multi-Radar Multi-Sensor (MRMS) product (*Smith et al., 2016*). The vertical coverage of MRMS as a function of location is given in **Figure 1**, which was created using the 3D reflectivity MRMS fields. Our region of interest for this study is the Continental United States (CONUS), east of the Rocky Mountains, over which radar beam blockage issues are minimal. As the radar beam propagates away from the transmitter it is progressively higher above Earth's surface due to both the curvature of the earth and the non-zero elevation angle of the beam itself (minimum of 0.5° for the operational Next-Generation Radar; NEXRAD). A comparison of REFC for Hurricane Dorian off the Florida coast with GOES observations indicated that when the vertical coverage falls below 70%, implying that only echoes above 3 km can be measured, the estimate of REFC becomes questionable. When only 50% of the vertical levels are present, this implies that only echoes above 6 km can be measured, and it appears that REFC provides very little reliable information. Over the Great Plains, where dew point depressions are large and cloud bases are higher than in the tropical environments of hurricanes, the reliability of REFC might fall-off with distance more slowly. In order to use the best quality radar data, we are restricting our domain of interest to east of 105°W, for which nearly all locations have 70% coverage, and most areas (by virtue of their population) have 90% coverage, or a minimum height of 1.25 km.

*d. Dataset Construction*

The first step in constructing a dataset for training ML is to resample all the inputs and outputs to a common grid. Since the goal of this work is to use the results for data assimilation, we have chosen the 3 km HRRR mass grid as the target grid. The projection and grid parameters are provided in **Table 1**, the formulae used for constructing the Lambert Conformal Conic and Cylindrical grids are given by *Synder (1987)*, and the formulae for the geostationary projection are provided by *Harris Corporation (2016)*. The MRMS grid is nominally 0.01° or roughly 1 km, and the GOES grid for the infrared bands used in this study is 2 km, so resampling to 3 km has minimal distortion. We note that due to averaging, after resampling MRMS to 3 km REFC values above 60 dBZ are very rare. The second step in preparing the data for training a CNN is to scale the inputs and outputs to the range 0-1. The scaling parameters for each variable are given in **Table 2**.

In order to reduce data volume and have the CNN focus on scenes of interest, Storm Prediction Center (SPC) filtered storm reports are used to automatically define regions and times of interest in order to maximize the number of storm reports (tornado, hail, wind). We selected samples from the 92-day period 4/17/2019 to 7/17/2019 during which there was abundant severe weather. The regions consisted of 256x256-pixels on 3-km HRRR grid (768x768 km) and 6-hour periods with 15-minute refresh. A histogram of the number of storm reports per day has a mode between 20-50 reports per case. Each case represents a 6-hour period on each day, which may span 0Z. **Figure 2a** shows that this construction approach results in a geographic preference for the Upland South and Southern Great Plains. **Figure 2b** shows a temporal preference for mid- to late-afternoon into the early evening. We split the data using a chronological 80% - 20% split for training-validation. Based on this split, the July cases were used for validation, while April-June was used for training. We have a total of 1798 samples for training and 448 samples for validation.

*e. Selection of Convolutional Neural Network Architecture*

This particular ML problem takes images as inputs and returns images as outputs, making this an image-to-image translation problem. The U-Net architecture is ideally suited (*Ronneberger et al. 2015*) to this problem type, and **Figure 3** shows the model we used. The model is drawn with optional skip connections, but for the results we will present we turned those connections off because they only provided small improvements and complicate the visualization (Section 3d). For this particular application, the GOES data provides mostly cloud-top information, while the radar provides information from deeper inside the cloud, thus the high-resolution spatial information that skip connections provide is not necessarily helpful.

The CNN depicted in **Figure 3** has three encoding and three decoding blocks. Each of the three encoding blocks consists of a convolution layer followed by a pooling layer. A pooling layer reduces resolution and allows the subsequent layers to detect patterns of larger spatial extent. Each decoder block consists of a convolution layer followed by an up-sampling layer. Up-sampling layers can be thought of as the (imperfect) inverse of a pooling layer, namely increasing resolution and using interpolation to generate an approximation. The convolutional filters are 3x3. While U-Nets often double the number of filters per convolution layer going down the encoding branch, and likewise halve the filters going up the decoding branch, we found this produced very small improvements. Instead, we used a constant number of filters, namely 32 filters/convolution layer. Using more than 32 filters/layer was unnecessary and would leave many inactivated. Using fewer filters/layer, such as 16, gave similar overall statistics as 32, but the outputs were noticeably blurrier.

As noted above, there are three encoder and three corresponding decoder layers. Based on an analysis of training and validation losses, we found that going deeper resulted in overfitting. Note also the choice of using only one convolution layer per encoder/decoder block, while U-Nets often use two convolution layers per block. Using two convolution layers per block doubles the number of trainable parameters, also making the chance of overfitting more likely. We are concerned with warm season convection, a phenomenon that is inherently small scale (e.g. meso-$\gamma$ to the smaller end of meso-$\alpha$), and a network of this depth and architecture performs well. However, for larger spatial phenomena, such as hurricanes and synoptic-scale frontal precipitation, a deeper network would likely be required. In such cases, more samples would be needed for training. When additional real samples are unavailable, data augmentation is the next best approach. As a side note, we found we could obtain similar results as those shown in this paper with a training dataset having 10x fewer samples by doing 10x augmentation, done by adding random noise to the real samples. However, the results shown herein used no data augmentation. Overall, training the GREMLIN model with 100 epochs yields the validation statistics: RMSD = 5.29 dBZ and $R^2$ = 0.738.

*f. Design of Loss Function to Address REFC Class Imbalance*

An important consideration in training the NN is the loss function, since radar reflectivity fields suffer from a class imbalance issue with an exponentially decreasing distribution for high values. In this section we discuss a new way to design a loss function to balance good performance for the rare (but important) high values with good performance for small values.

Training the NN using the standard unweighted mean-square-error MSE loss function results in sub-optimal performance at high REFC (**Figure 4**). High radar reflectivity values are relatively less common: if *y* represents the scaled radar reflectivity (scaling 0-60 dBZ linearly into the range 0-1), then the probability density function is closely approximated by $P(y) \propto e^{-5y}$ with an $R^2$=0.80. We developed a novel method of using a performance diagram (**Figure 4**) to select loss function weights that produce the minimum categorical bias. Categorical statistics are discussed in *Wilks (2006)*, and the binary categories are created by evaluating whether the true and predicted REFC are greater than a threshold. While minimizing the categorical bias does not guarantee that the results will also have maximal critical success index, we found that in practice this was in fact the case.

Our approach is related to using an Area Under the Receiver Operating Characteristic Curve as a loss function but avoids the problem of derivatives not existing for a discontinuous function. The approach also acts as a global constraint on the realism of the resulting fields by balancing overprediction and underprediction of reflectivity values across the spectrum. We define weights (*Wt*) for the MSE loss function (*L*) according to a generalized exponential:

$$L(y_{true}, y_{pred}) = \frac{1}{N}\sum_{j=1}^{N} Wt(y_{true})(y_{pred} - y_{true})^2 \qquad (2a)$$

$$Wt(y_{true}) = e^{by_{true}^c} \qquad (2b)$$

where *y*<sub>*true*</sub> and *y*<sub>*pred*</sub> are the true and predicted values of *y* and *N* is the number of training samples. We then vary *b* and *c* in a grid search to find the optimal values. Values of the categorical bias are calculated at each REFC threshold *i* from 5 to 50 dBZ in steps of 5 dBZ, and best matching model is found taking the parameter combination *k* with:

$$\min_{k}\left(\operatorname*{mean}_{i}(|1 - Bias_{i,k}|)\right) \qquad (3)$$

In order to get reliable results, we also train several versions of the model (20 versions) that differ only in their random seeds.

While the intuitive 1/PDF weights would give *b*=5 and *c*=1, we found the minimum categorical bias weights were *b*=5 and *c*=4 for the MSE (mean-square-error) loss and *b*=5 and *c*=3 for the MAE (mean-absolute-error) loss. The disparity suggests there might be a way to choose coefficients from first principals based on the PDF, but we note that the best results require a much heavier weighting of the high values than would be implied by direct usage of the inverse of the PDF. Convergence plots for categorical statistics tracking progress during NN training show that performance converges quickly at low REFC values that are abundant and are noisier and take longer to converge at high REFC values. Thus, it is crucial to train the NN long enough; here, we used 100 epochs. We denote the trained CNN with architecture and loss function as described in this section as GREMLIN Version 1.

## 3. Results and Discussion
*a. Baseline Network Performance*

The overall performance of our final neural network, GREMLIN, is shown as the red line in Fig. 4. To understand the abilities of GREMLIN to produce synthetic radar reflectivity, it is helpful to consider a specific example. **Figure 5** compares MRMS REFC (a,c,e) with GREMLIN -derived REFC (b,d,f) at three times during the event (21Z, 23Z, 01Z), noting that the first large hail reports were at 20:50Z and lasted until 21:30Z. This case is notable because of its severe impact on the Denver Metropolitan Area; the storms produced up to baseball sized hail (2.75 inches) and was the costliest weather catastrophe in Colorado – producing $1.4 billion in insured losses (*Svaldi, 2017*). In addition to its human impact, this case poses challenges for both infrared imagers and optically sensed lightning. It is an example of Great Plains thunderstorms with abundant cloud water concentrations (e.g, *Williams et al., 2005*) that produce large anvils that obscure the convective cores in infrared imagery. While these conditions also lead to very high lightning rates, *Rutledge et al. (2020)* show these conditions also produce storms for which the lighting flash height is relatively low, making for large optical paths between the lightning source and the upper cloud boundary along the GLM sensor line of sight (both in general and for this particular case). This regionally common "inverted" charge structure causes a relative minimum in lightning detection efficiency over the Great Plains (*Marchand et al., 2019*).

Despite the challenges, **Figure 5** shows that GREMLIN performs well for this case. In the early stages (**Fig. 5a,b**) GREMLIN captures the three distinct convective cores near Denver, Greeley, and Fort Morgan. It correctly represented the location of the strongest echoes, although it also tended to overestimate them, and the fine-scale structure of the cores is not captured. Two hours later (**Fig. 5c,d**) as the storms began to transition to a convective line morphology, the GREMLIN estimates captured that transition well. GREMLIN properly located the strong echoes, although small areas that were distinct in MRMS tended to get merged in GREMLIN. Finally, after dark (**Fig. 5e,f**) and as the convection transitioned from distinct cells to lines, GREMLIN captured the basic shape and curvature of the lines. While it merged the two lines in northern Colorado, it kept the strong echoes over southern Colorado distinct.

Characterizing the spatio-temporal performance of the technique is complicated by the natural variation of convective morphology. In our training dataset, convection tends to be more widespread in the eastern U.S., while isolated convective cells are more common in the west. Since RMSD statistics are sensitive to the echo coverage fraction $F$, care must be taken to separate true regional biases from artificial biases that arise from natural regional variations in these properties. **Figure 6a** shows the RMSD versus the echo coverage fraction, $F$, (defined here by the 20 dBZ radar reflectivity contour). It can be seen that small $F$ are associated with small RMSD. It also shows that eastern U.S. regions tend to have both larger $F$ and larger RMSD. However, the easternmost locations do have errors greater than the average (black line given by RMSD = 2.2 $F^{0.36}$). Given that our training samples have a fairly uniform distribution from east-to-west (**Fig. 2a**), the fact that the predictions exhibit an "Oklahoma-centric" bias is notable and may be a consequence of using a loss function that is heavily weighted toward higher REFC values.

The typical lifecycle is for convection to initiate with the heating of the daytime, and then grow upscale overnight. One might expect the large echo structures at night to validate better since the GREMLIN estimates tend to be more smoothed out than MRMS REFC. To look for biases in time, **Fig. 6b** gives the RMSD vs $F$ as a function of the solar zenith angle, where sunset is 90°. It does show a population of samples that have both large $F$ and small RMSD, however most nighttime samples are below the average line, even at smaller $F$. This good performance at night is notable given that our training samples emphasize late afternoon and early evening (**Fig. 2b**). It is possible this is a result of GLM having a 20% higher detection efficiency at night than during the day (*Marchand et al., 2019*). Not all daytime retrievals have lower skill, and the day/night distinction in skill is less clean than the east/west distinction. However, since daytime retrievals do have room for improvement, this argues that the solar reflective bands, visible and cloud particle phase/size bands in particular, should be used.

Overall, GREMLIN performs well. In particular, GREMLIN is able to accurately locate areas of strong echoes, which have been difficult to capture with heritage methods (e.g., *Arkin and Meisner, 1987*).

*b. Targeted Architecture Experiments*

A key question raised by the results shown in Section 3a is: "*What is the network learning to produce such good skill?*" We use several different methods to answer this question, starting with targeted architecture experiments. Namely, we modify the GREMLIN architecture by removing specific capabilities. Analyzing the performance of the resulting restricted NNs tells us which capabilities of GREMLIN are most essential for its success and sheds light on how they are used.

We begin by removing the capability of GREMLIN to utilize information on radiance gradients and spatial context used by the network - done by replacing all 3x3 filters by 1x1 filters. Secondly, we trained models withholding sets of channels. **Figure 7** provides results for a representative validation sample. For simplicity, we focus on the impact of gradients in Channel 13, which is the most important channel (Section 3d), and on lightning information. The C13 $T_B$s (**Fig. 7a**) exhibit very sharp spatial gradients from clear areas with $T_B > 275$ K to areas with radar echo with $T_B \sim 220$ K. Comparing with the spatial pattern of REFC (**Fig. 7c**) it can be seen that cold $T_B$s are generally a good predictor that a particular pixel has REFC > 15 dBZ, but there is a low spatial correlation between the coldest $T_B < 215$ K and the higher REFC values > 35 dBZ. These areas of strong echoes correlate well with lightning (**Fig. 7b**), although the lightning is a bit smoother than REFC and there are spatial displacements. The latter may be due to a combination of residual parallax displacement errors and the effects of vertical wind shear.

**Fig. 7d-i** shows the progression of results for six NN models with increasing capabilities, from the most restricted model (**Fig. 7d**) to the full model, GREMLIN (**Fig. 7i**). The 1x1 filter experiments are shown in the middle row (**Fig. 7d,e,f**), which represents the performance that could be expected from a traditional pixel-based retrieval. With C13-alone (**Fig. 7d**) the areas of REFC > 15 dBZ are reasonably well delineated, but it completely lacks any echoes > 35 dBZ. Combining GLM with C13 (**Fig. 7e**) shows huge improvements in the representation of echoes > 35 dBZ, although the spatial extent is a bit too large. Bringing in the other two channels (C07 and C09) in **Fig. 7f** does help reduce the errors a bit. So, without the use of spatial gradient information, lightning information is critical to obtaining any skill for higher REFC values.

The bottom row (**Fig. 7g,h,i**) shows the results using 3x3 filters. Even with C13-alone, the use of gradient information and spatial context (**Fig. 7g**), produces marked improvements in skill, especially at the high REFC end. Compared with the 1x1 experiment (**Fig. 7d**) the probability of detection (POD) jumps from 0 to 0.24, and the false alarm rate (FAR) of 0.59 is slightly better than using all channels with no spatial information (**Fig. 7f**). Adding lightning information (**Fig. 7h**) more than doubles the POD and also reduces the FAR. Adding the other channels (**Fig. 7i**) helps as well, producing significant improvements in RMSD and $R^2$, also resulting in higher POD and lower FAR. We hypothesize that results of this quality (**Fig. 7i**) are sufficiently good to produce a positive impact on data assimilation.

The results for this example are consistent with those across all validation samples (**Fig. 8**). Without the benefit of spatial information and lightning (black and green lines in **Fig. 8a**), the RMSD at high REFC is as large as 25 dBZ. Note that removing spatial context, but adding lightning (blue line **Fig. 8a**), makes the RMSD slightly worse for REFC in the range 20-35 dBZ, but produces large improvements above 35 dBZ, bringing the RMSD down to 15 dBZ. Adding spatial context yields additional large improvements (**Fig. 8d**). Combining spatial information and lightning produces the best results, with RMSD of 12 dBZ at the highest REFC. Without spatial information, lightning shows obvious value in increasing the POD (**Fig. 8b**) and reducing the FAR (**Fig. 8c**). In the absence of lightning information, adding the water vapor channel (green line **Fig. 8b,c**) does provide some improvements in POD and FAR, but not as much as lightning.

Based on examining predictions, it appears the network correlates smaller differences between C13 and C09 with higher REFC. However, those areas of small C13-C09 difference tend to be more spatially extensive than REFC, with the result being that POD is improved, but FAR is slightly worse. This finding demonstrates the unique benefits of lightning information to pinpoint the areas of strong updrafts and high REFC. When spatial information is used, the value of lightning is relatively less, but it still makes significant improvements in POD (**Fig. 8e**) and FAR (**Fig. 8f**). Further insights into how the network is using lightning and spatial information together is provided by use of attribution methods (Section 3d).

*c. Examining the Effective Receptive Field*

GREMLIN is a *purely* convolutional neural network, i.e. it does not have any fully connected (aka dense) layers. This means that any individual output neuron, i.e. any pixel of the estimated MRMS image, is connected to only a small group of input neurons corresponding to a small spatial neighborhood of the output pixel in the input channels. This small area is known as a CNN's *Receptive Field* (*Luo et al. 2016*). For our application the receptive field tells us the maximal spatial context size and thus the maximal size of a meteorological feature that can be recognized and utilized by GREMLIN to determine the value of a single pixel of the estimated MRMS image.

One can calculate the maximal extent of the receptive field from the CNN architecture (*Araujo et al., 2019*). However, pixels at the center of the receptive field have the largest impact, with impact decreasing rapidly for pixels further away in a roughly Gaussian distribution (*Luo et al 2016*). It is useful to view the actual distribution of the receptive field, the *Effective Receptive Field* (ERF, *Luo et al 2016*), to understand which size neighborhood truly has a significant impact. The ERF, which depends on the network's weights, changes during training. Thus, it cannot be calculated from architecture alone. Here, we develop an ERF approximation based on the SmoothGrad algorithm (*Smilkov et al., 2017*). The approximation is described in detail in Appendix A.

**Figure 9** shows our approximation of ERF for GREMLIN for different lengths of training, ranging from an untrained model with random weights (a) to the final model trained for 100 epochs (d). Each ERF image in Fig. 9 shows the cumulative results across all four channels.

Note that the ERF consistently occupies a region of less than 53x53 pixels (red squares in **Fig. 9**) with the region of highest impact actually much smaller than that, especially in the trained models. The ERF of the untrained model is the most spread out (a). Early training (b,c) seems to make the model put more emphasis toward the center, potentially as a sort of first-order approximation. The final model retains some focus in the center, but also spreads out more– potentially moving beyond the first-order approximation and taking additional detail into account. While the results in **Fig. 9** are only ERF approximations (details in Appendix A), and vary across considered samples, output pixels, and random seeds used to train the CNN, we conducted many more experiments and found the trends in **Fig. 9** to be representative of the overall behavior of the ERF distributions. Please see the detailed comments in Appendix A on the interpretation of such ERF approximations.

*d. Applying Attribution Methods to Identify NN Strategies*

To learn more about the underlying logic GREMLIN uses to derive its estimates, we use the method of layer-wise relevance propagation (LRP). Given an input sample and an output pixel, LRP tells where the neural network was primarily looking when deriving the output pixel's estimate. We find that LRP is better suited for this purpose than standard gradient-based methods because LRP takes a global view of this decision-making process, rather than just taking a local derivative as gradient-based methods do. Details of LRP are provided in Appendix B.

**Figure 10** shows LRP results for GREMLIN for the same sample as in **Fig. 9**, but in this case focusing on a different output pixel, chosen for its close proximity to strong lightning activity. All panels in **Fig. 10** are zoomed into a neighborhood of the chosen output pixel. The first row shows the input channels and corresponding desired output (i.e., the MRMS observations). Because we suspected that the neural network was heavily reliant upon the gradient of the input channels, we show an approximation of the input channel gradient magnitudes in the second row. These gradient magnitudes were calculated by applying a Sobel operator (*Gonzalez and Woods, 2002*) to the input channels. The gradient estimates are not fed into the neural network; they are provided here simply to highlight the locations of the strongest gradients. The third row of **Fig. 10** shows the first set of results, namely the LRP maps of where in the input channels the neural network pays attention in order to estimate the value of the chosen output pixel for this sample, along with the estimated MRMS results.

The LRP result for the GLM channel shows that the NN focused only on regions where lightning was present in that channel. The LRP results for the other channels show that even in those channels the NN's attention was drawn to focus on regions where lightning was present. We then performed a new experiment by modifying the input sample to have all lightning removed, i.e. the GLM channel was set to all zero values. For this case LRP showed us that the network's focus shifted entirely to the first three input channels, as expected. More importantly, the focus shifted to the *gradient* of the input channels, as can be seen by comparing the three left-most panels of the second and fourth row. In fact, near the center of the fourth-row panels, it is striking how similar the LRP patterns of those three channels are to the strongest gradient lines in the second row. LRP vanishes further away from the center location, as expected given the nature of the ERF properties.

These results indicate the following strategy used by GREMLIN: *whenever lightning is present near the output pixel, the NN primarily focuses on the values of input pixels where lightning is present, not only in the GLM channel, but in all four input channels.* It seems that the network has learned that locations containing lightning are good indicators of MRMS behavior, even in the other input channels. *In the absence of any lightning, the NN focuses on locations where the gradient is strong.* It seems to have learned that those locations have the highest predictive power for estimating the output. Additional experiments confirmed these two strategies of the final neural network for a wide selection of samples and output pixels.

*e. Synthetic Inputs to Quantify Sensitivity to Radiance Gradients*

The use of architecture experiments (Section 3b) and attribution methods (Section 3d) have demonstrated the importance of radiance gradients for retrieving high REFC values. In this section, we construct synthetic inputs and probe the network's response to quantify that sensitivity. For this purpose, we enlist a sum of Generalized Elliptical Gaussians (GEG) model. This model assumes an outer Gaussian ($G_o$) that represents the thunderstorm anvil, and an inner Gaussian ($G_i$) that represents the overshooting top. The synthetic brightness temperature ($T$) is a function of ($x,y$) with the following parameters: location $x_0$ and $y_0$, amplitude $A$, size $S$, aspect $\alpha$, orientation $\theta$, and sharpness (exponent) $p$ for the outer and inner Gaussians, denoted with subscripts $o$ and $i$:

$$\hat{x}_{o,i} = (x - x_{0,o,i}) \cos \theta_{o,i} - (y - y_{0,o,i}) \sin \theta_{o,i} \tag{4a}$$

$$\hat{y}_{o,i} = (x - x_{0,o,i}) \sin \theta_{o,i} + (y - y_{0,o,i}) \cos \theta_{o,i} \tag{4b}$$

$$T_{o,i} = exp\left(-1\left(\frac{\hat{x}_{o,i}^2}{2S_{o,i}^2} + \frac{\hat{y}_{o,i}^2}{2(S_{o,i}\alpha_{o,i})^2}\right)^{p_{o,i}}\right) \tag{4c}$$

$$T = A_o T_o + A_i T_i \tag{4d}$$

Evaluating thousands of different parameter settings, the spatial patterns that most strongly activates the network, based on the maximum REFC, all resemble **Fig. 11a**. What the strongly activating patterns have in common, and what is different from the weakly activating patterns, are very large $p_o$ and large $p_i$, meaning that the anvil and overshooting top have very sharp $T_B$ gradients. The other traits the strongly activating patterns have in common are that $G_i$ is located near the edge of $G_o$ and that $S_i \ll S_o$. The patterns producing a weak response tend to look unphysical from a meteorological perspective, indicating that the network has learned about realistic looking overshooting top signatures. This is a desirable property: rather than responding strongly to unphysical outlier inputs, it only responds strongly to patterns that look meteorological, although that does not rule out the possibility that the network could be fooled by a cleverly constructed counterexample. Out of all the parameters of the GEG model, the ones that are most influential in producing high REFC values are $p_o$ and $p_i$, and **Fig. 11b** characterizes the maximum REFC as a function of those parameters. The emergence of 35 dBZ echoes requires $p_o$ to be 1 or greater, or very large $p_i$ around 8. Thus, the CNN does not just respond to gradients, but calibrates its response based on the sharpness of the brightness temperature gradient.

## 4. Summary and Conclusions

This paper trained and evaluated a CNN that uses ABI infrared channels and GLM lightning data to estimate MRMS REFC over eastern CONUS during the warm season. Since REFC follows an exponentially decreasing distribution, to get good performance at high values, we used a weighted loss function. A variety of approaches were examined to investigate what the network learned and how it makes its predictions. Channel withholding experiments showed that geostationary lightning observations are uniquely valuable for their ability to pinpoint locations of strong updrafts. Experiments withholding spatial information demonstrated that radiance gradients carry more information about high REFC values than the radiance values themselves. Layer-wise relevance propagation established that the CNN uses the information from ABI and GLM in a synergistic manner, where it interprets ABI radiance gradients in the context of whether GLM indicates the presence of lightning. Synthetic input experiments confirmed that the sharper the gradient, the stronger the CNN response, but only for patterns that have an appearance reminiscent of meteorological convection.

Having established that the horizontal spatial patterns of radar reflectivity can be accurately estimated using GOES data, the next step in this research is to produce full 3D profiles of radar reflectivity for use as an input to data assimilation systems. Here, we may leverage ongoing research to estimate cloud geometric thickness (*Noh et al., 2017*) and vertical structure (*Miller et al., 2014*) via empirically-based methods. The current non-variational technique for initializing RAP/HRRR with radar reflectivity does not require characterization of uncertainty, however uncertainty information is required for variational approaches.

Over CONUS the results are easy to validate using retrospective simulation experiments where the actual radar data are withheld and replaced by the GOES estimates. However, the real value of the technique will come from its ability to fill in locations that lack radar coverage due to terrain blockage, which are mostly over the western U.S. and coastal/oceanic locations. Evaluating results in these locations is much more difficult due to a lack of observations. However, MRMS sectors over the Caribbean (GOES-16), Hawaii (GOES-17), and Guam (Himawari-8) do provide observations, as do spaceborne radar reflectivity observations from the Global Precipitation Measurement (GPM) Dual-frequency Precipitation Radar (DPR). How well the model derived in this paper will generalize to meteorological regimes outside of the training set is an open question. However, it is known that both lightning and storm characteristics are different over land versus ocean *(Nag and Cummins, 2017; Bang and Zipser, 2015)*. Thus, additional contextual information that is geographic or meteorological in nature may be needed, along with a deeper network to accurately depict features at the upper end of meso-α to synoptic scales.

**Data Availability Statement**

This study uses publicly available datasets. The L1b ABI data files used in this study are available from NOAA CLASS: https://www.bou.class.noaa.gov/saa/products/search?datatype_family=GRABIPRD.
The L2 GLM data files used in this study are available from NOAA CLASS: https://www.avl.class.noaa.gov/saa/products/search?datatype_family=GRGLMPROD.
The MRMS composite reflectivity data files are available from NCEP: https://mrms.ncep.noaa.gov/data/.

**Acknowledgements**

We would like to thank the GOES-R Program for supporting this research with award NA19OAR4320073 and the National Science Foundation for supporting this research through grant HDR-1934668 (Ebert-Uphoff).  We would like to thank NOAA RDHPCS for access to the Fine Grain Architecture System on Hera, without which this research would not have been possible.

**Appendix A: Method for Approximating the ERF**

To get an estimate of the ERF we want to calculate and visualize how much each location in the input channels affects a specific output pixel in a considered neural network. A simple way to do so for a given input sample and chosen output pixel is to calculate the gradient of the output neuron with respect to the neurons in the input channels. Calculating this gradient is a common task in neural networks and built-in routines are readily available in neural network computing environments. However, the results tend to be noisy and we thus use a modification of this approach, namely the SmoothGrad algorithm by *Smilkov et al. (2017)*. SmoothGrad calculates the gradient with respect to the input neurons several times, each time adding Gaussian noise to each pixel of each input channel before calculating the gradient, and then returns the average result. This approach, as the title of the *Smilkov et al. (2017)* aptly states, removes noise (in the results) by adding noise (in the input channels).

We use the SmoothGrad implementation of the "tf-explain" package (see https://tf-explain.readthedocs.io/en/latest/) with 100 samples and a noise level of 1.0. Note that this noise level is chosen extremely large on purpose (keep in mind that our inputs are scaled to values between just 0 and 1), because that makes the results less dependent on the specific sample that was chosen for the estimation. When interpreting the resulting ERF estimates for a neural network model one should keep in mind that the results vary based on i) chosen input sample, ii) chosen output pixel, and iii) random noise generated by SmoothGrad. Thus, it is important to generate estimates for variations of all these parameters and ensure that results are representative of the general trends. A property we noticed varying across those parameters is the presence of a few high intensity pixels in the resulting maps. Their number and location can vary and thus should not be assigned special meaning. Aside from such details the overall distribution is fairly consistent, namely how diffuse the ERF is and how far it stretches out from the center. More generally, *results from this ERF approximation method should be seen as a random sample drawn from a given distribution, rather than each pixel value given specific meaning.*

**Appendix B: Layer-wise Relevance Propagation (LRP)**

A key idea of layer-wise relevance propagation is that it seeks to track *relevance* backward from an output neuron to the input image, by tracking backwards which neurons in the prior layer were most responsible for the values of a neuron in the later layer. To do so LRP does not use any of the built-in backpropagation rules of neural networks and develops instead its own set of customized rules. By applying those rules iteratively, an overall estimate of relevance in the input space is obtained. LRP is a fairly complex topic and the details are beyond the scope of this paper. For a detailed introduction see *Bach et al. (2015)*, *Montavon et al. (2018)*, or *Toms et al. (2019)*.

We are using the implementation of LRP in the "innvestigate" package for Tensorflow (see https://innvestigate.readthedocs.io/en/latest/). We are using the alpha-beta rule (Eq. (60) in *Bach et al. (2015)*) with alpha=1 and beta=0, to only approximate positive attribution, i.e. to identify locations for which higher activation values tend to make high values at the output *more* likely. We had to use a few tricks to make this implementation work for our purpose. Firstly, we flattened the output layer of the NN into a vector to be able to prescribe which output pixel we want to look at. Secondly, we did not use the standard heatmap visualization provided by the package, but instead split the heatmap result for LRP into its separate channels and plotted them separately. *For the interpretation of LRP results one needs to keep in mind that LRP uses approximation rules and that it was specifically designed for classification tasks, not regression tasks, so results should always be interpreted as showing overall trends but should not be interpreted on a pixel-by-pixel level.*

**Tables**

**Table 1.** Projection and grid parameters for each dataset.

| GOES | | MRMS | | HRRR | |
|---|---|---|---|---|---|
| **Parameter** | **Value** | **Parameter** | **Value** | **Parameter** | **Value** |
| Projection | Geostationary | Projection | Cylindrical | Projection | Lambert Conformal Conic |
| Altitude | 35786023.0 m | Lower left longitude | -130ºE | Reference longitude | 262.5ºE |
| Equatorial radius | 6378137.0 m | Lower left latitude | 20ºN | Reference latitude | 38.5ºN |
| Polar radius | 6356752.31414 m | Longitude scale | 0.01 | Standard parallel | 38.5ºN |
| Center longitude | -75.0ºE | Longitude dimension | 7000 | X scale | 3.0 km |
| X scale | 5.6e-05 | Latitude scale | 0.01 | X dimension | 1799 |
| X offset | -0.101332 | Latitude dimension | 3500 | Y scale | 3.0 km |
| X dimension | 2500 | | | Y dimension | 1059 |
| Y scale | -5.6e-05 | | | Earth radius | 6370 km |
| Y offset | 0.128212 | | | | |
| Y dimension | 1500 | | | | |

**Table 2.** Scaling parameters for each variable, each scaling is linear.

| Channel | Minimum | Maximum | Inverted |
|---|---|---|---|
| **C07** | 200 K | 300 K | True |
| **C09** | 200 K | 250 K | True |
| **C13** | 200 K | 300 K | True |
| **GLM** | 0.1 groups 5-min$^{-1}$ km$^{-2}$ | 50 groups 5-min$^{-1}$ km$^{-2}$ | False |
| **MRMS** | 0 dBZ | 60 dBZ | False |

**Figures**

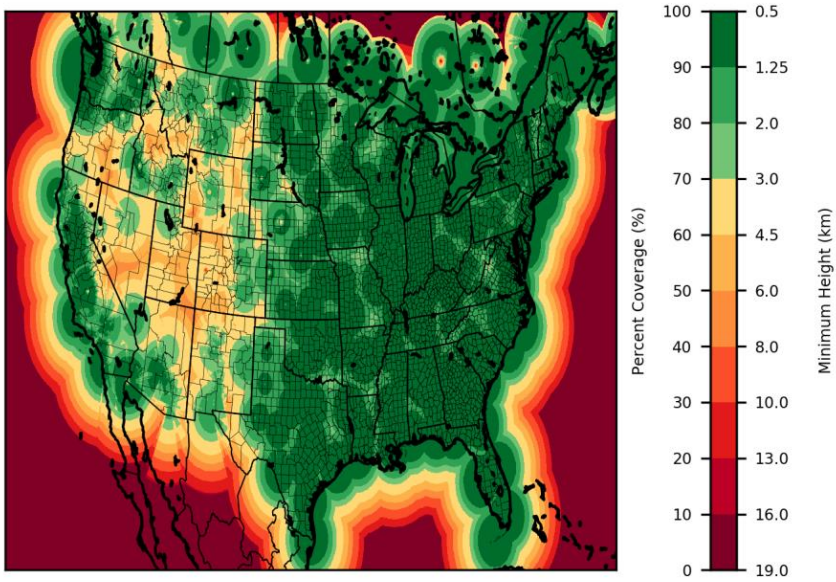

**Figure 1.** MRMS radar coverage in terms of the percent of MRMS levels available at each location and the minimum height at that location.

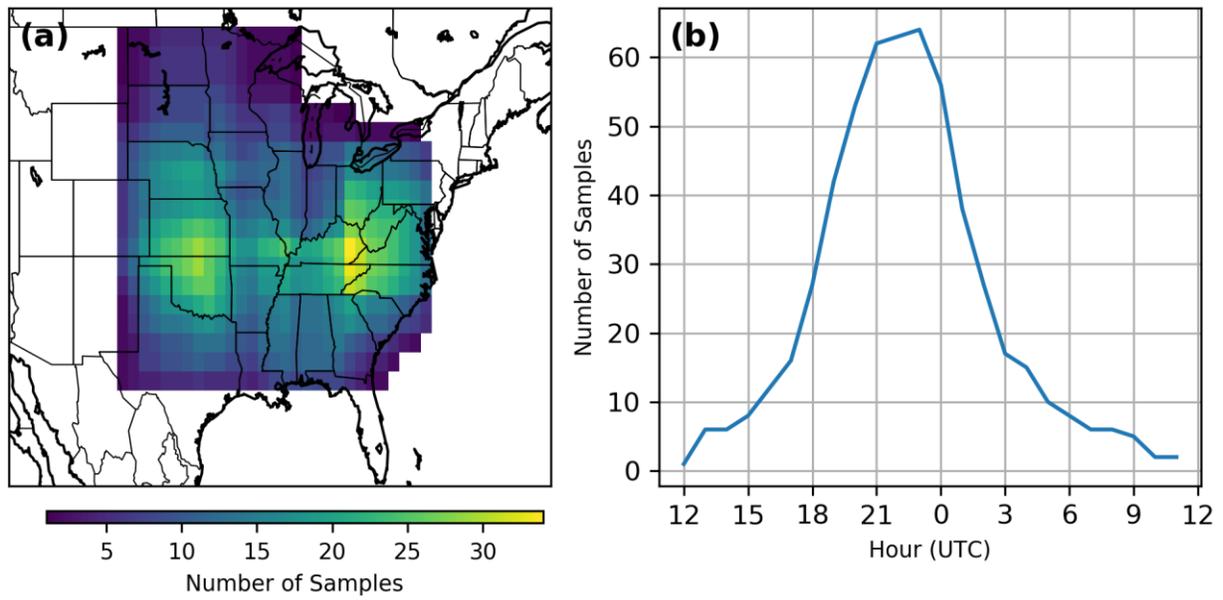
**Figure 2.** (a) Spatial distribution of samples. (b) Temporal distribution of samples.

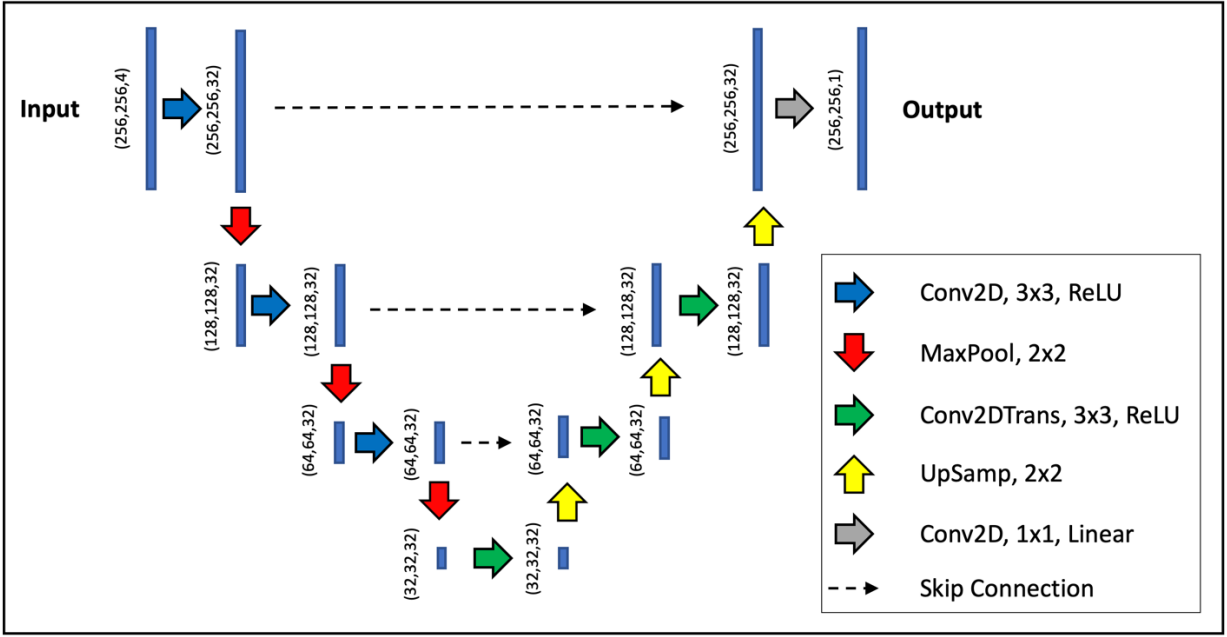

**Figure 3.** U-Net architecture for a model with 47,457 trainable parameters.

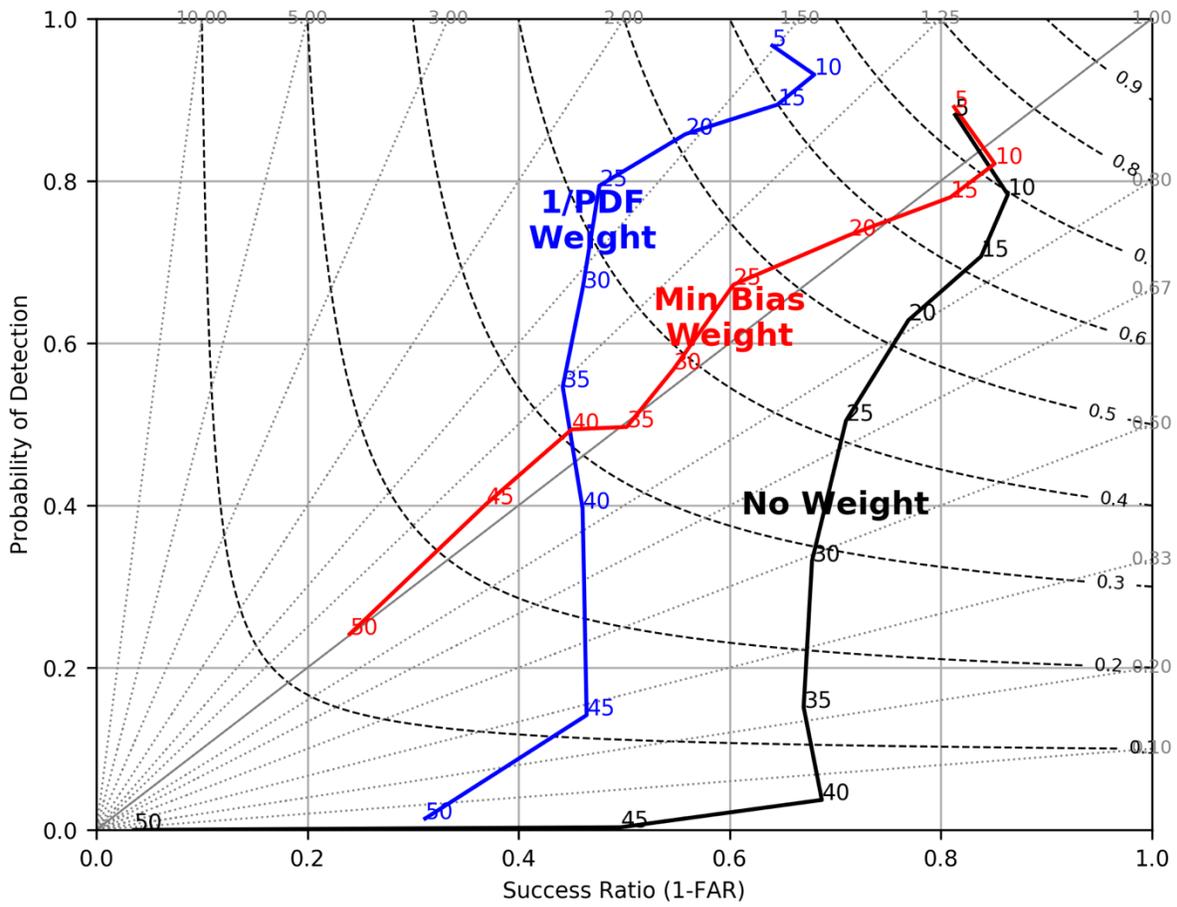

**Figure 4.** Performance diagram for REFC categories 5, 10, …, 50 dBZ. Dashed black contours are critical success index, and grey dotted lines are categorical bias. The solid black line is performance using unweighted MSE loss function, solid blue line uses 1/PDF weighted MSE loss function, and the solid red line uses weights that produce the minimum categorical bias (GREMLIN).

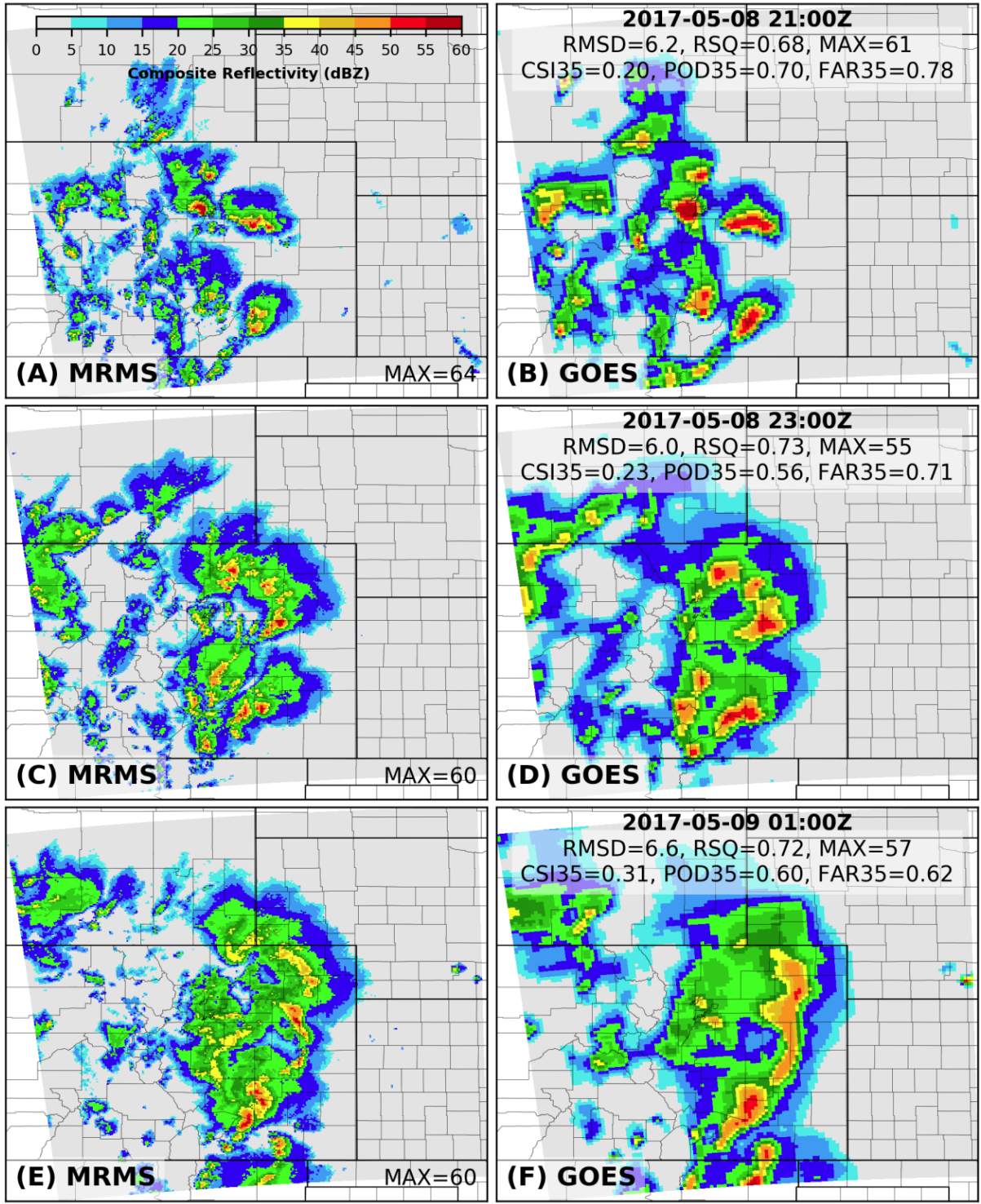

**Figure 5.** Colorado 2017-05-08 case: MRMS (a), (c), (e); GREMLIN prediction (b), (d), (f).

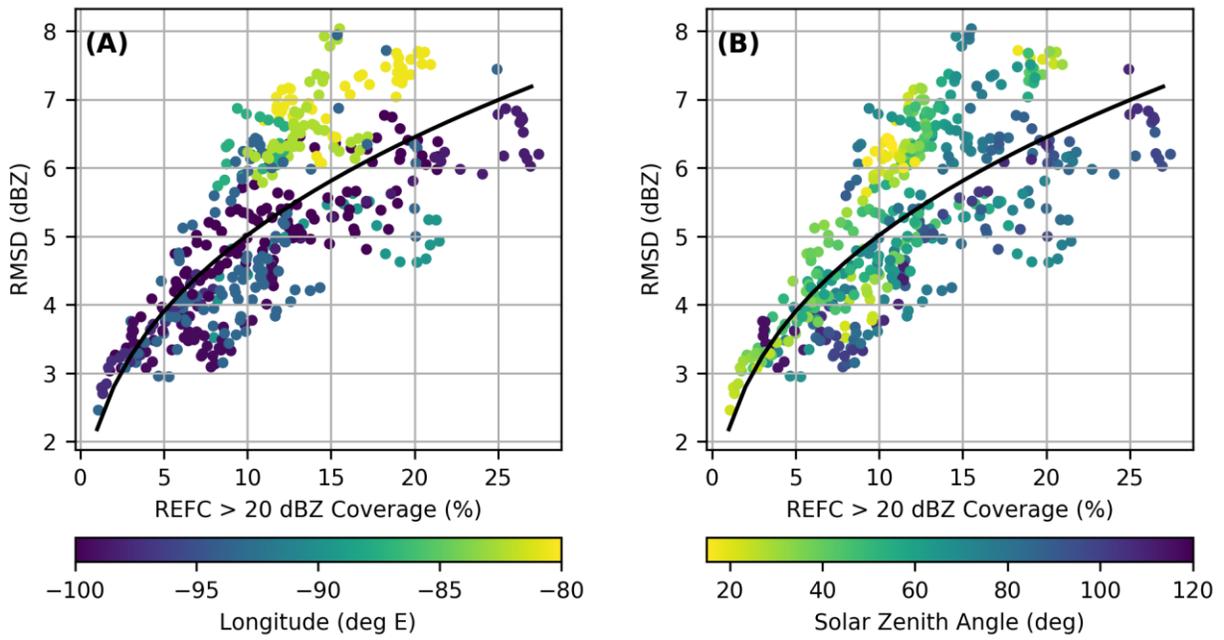

**Figure 6.** RMSD versus the percentage coverage of radar echoes > 20 dBZ where the color indicates: (a) longitude and (b) solar zenith angle.

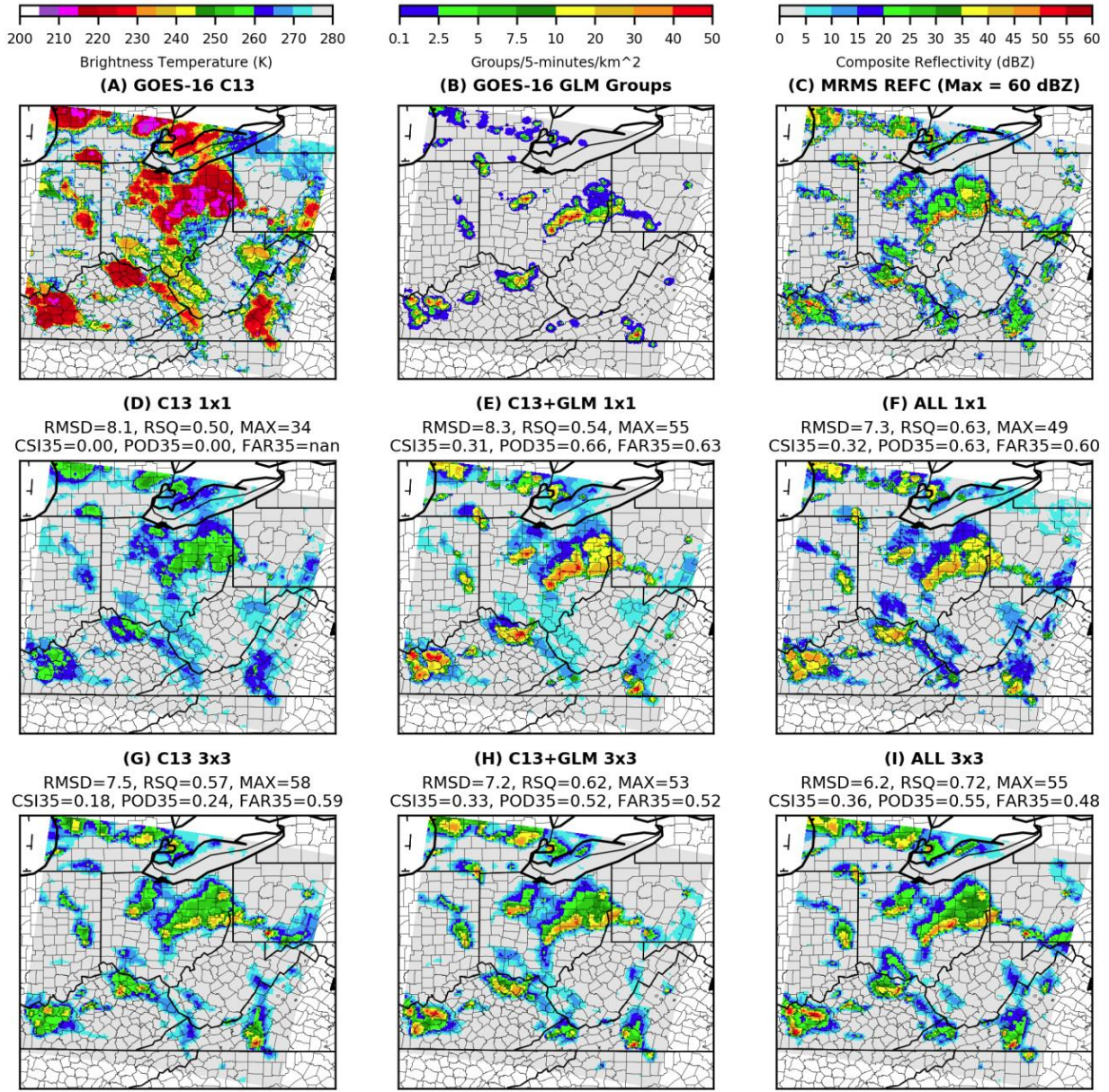

**Figure 7.** Validation sample 2019-07-02 23:30Z inputs: (a) GOES C13 and (b) GOES GLM; truth: (c) MRMS; and prediction for progression of six models with increasing capabilities, (d) 1x1 filters C13-only, (e) 1x1 filters C13+GLM, (f) 1x1 filters all channels, (g) 3x3 filters C13-only, (h) 3x3 filters C13+GLM, (i) 3x3 filters all channels (GREMLIN). Panels (d)-(i) provide the following statistics: root-mean-squared-difference (dBZ), coefficient of determination, maximum REFC (dBZ), 35-dBZ critical success index, 35-dBZ probability of detection and 35-dBZ false alarm rate.

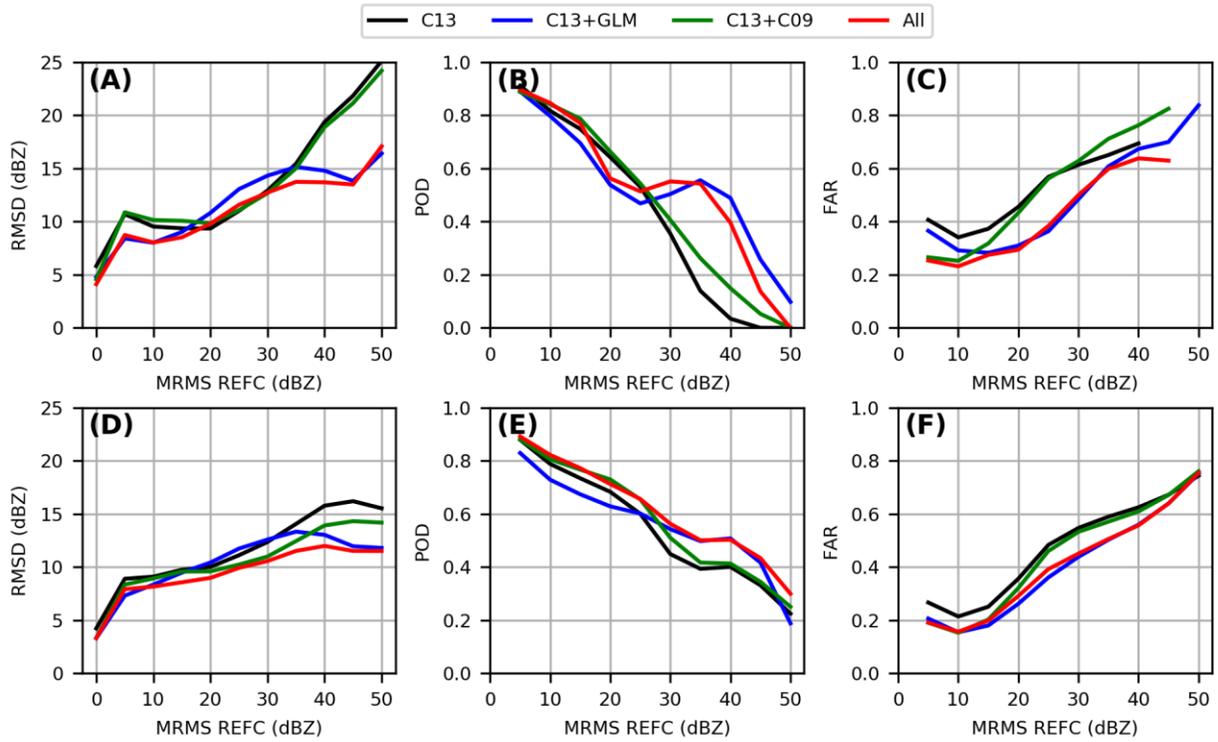

**Figure 8.** Statistics for 1x1 filters: (a) RMSD, (b) POD, and (c) FAR vs REFC for various experiments (line colors). Statistics for 3x3 filters: (d) RMSD, (e) POD, and (f) FAR vs REFC for various experiments (line colors).

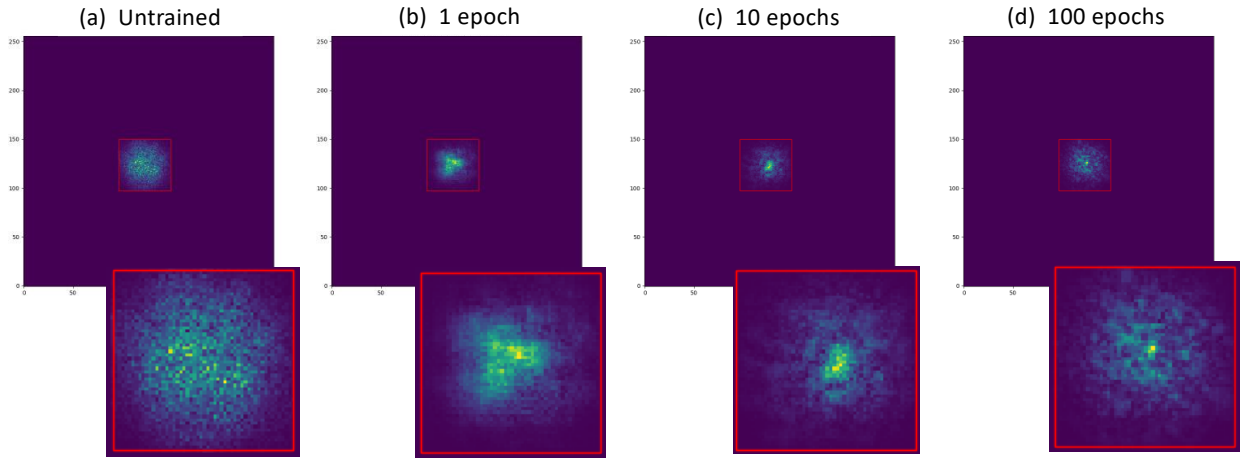

**Figure 9.** ERF approximation for four different models with identical architecture (architecture of GREMLIN), but different lengths of training, ranging from no training (a) to fully trained model, GREMLIN (d). For each image we show the ERF in the original 256x256 space of the input channels and a zoom-in of a 53x53 region (red box). Results are for Sample 68 and output pixel (125,125). (Note that the four models did not start out with the same random seed, thus cannot strictly be seen as a progression of training toward the final model, but rather as independently trained models with different training length.)

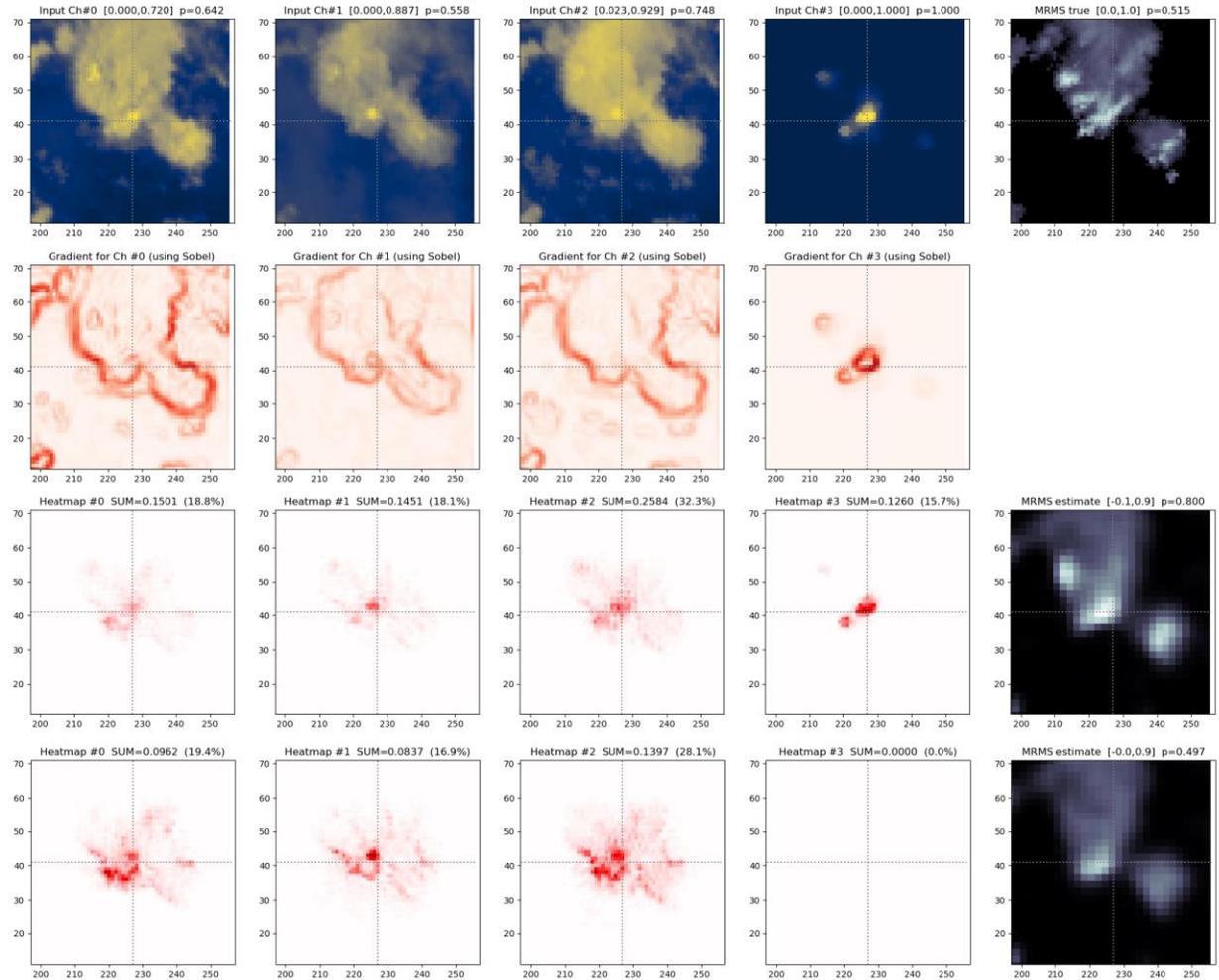

**Figure 10.** LRP results for GREMLIN for Sample 68 and output pixel (227,41). Top row shows the four input channels (left-to-right: ABI C07, ABI C09, ABI C13, GLM Groups) and the corresponding MRMS image (true values). Second row shows the gradient of the input channels calculated by applying a Sobel operator. Third row shows LRP results for the original four input channels and the chosen output pixel, and the MRMS estimate. The fourth row shows the equivalent of the third row, but after all values of the GLM channel were set to zero. Note that all images are zoomed in to a region centered at the pixel of interest.

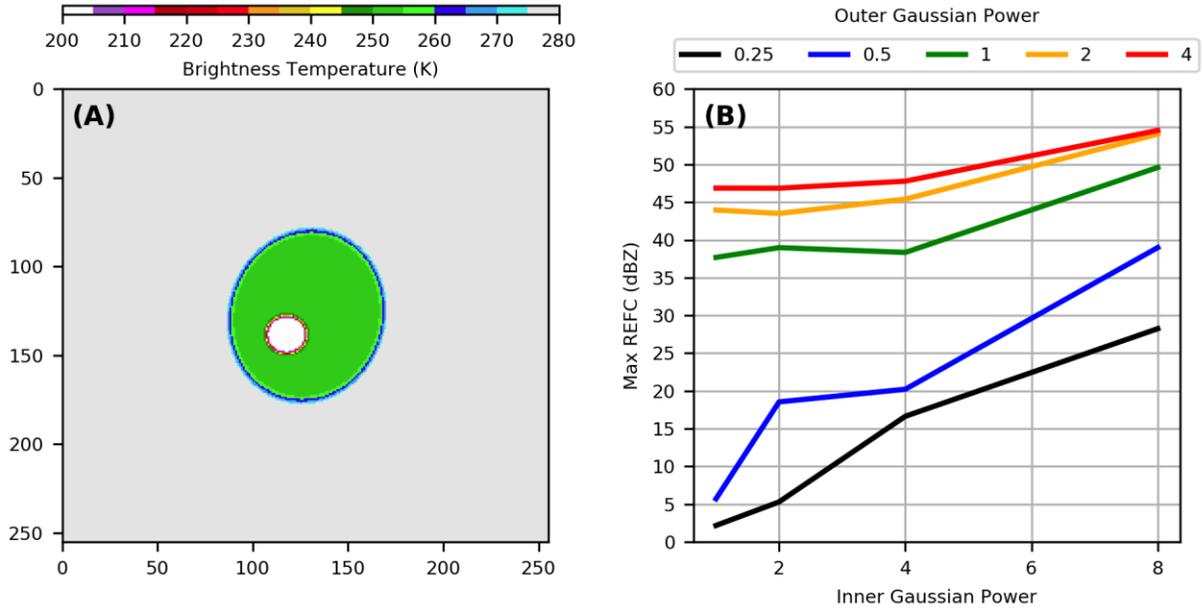

**Figure 11.** (a) Synthetic C13 $T_B$ that produces the maximum REFC response for GREMLIN. (b) Maximum REFC as a function of inner Gaussian power (x-axis) and outer Gaussian power (line color).